\documentclass[review]{elsarticle}
\usepackage{amsmath}
\usepackage{graphicx,amssymb}
\usepackage{array}
\usepackage{multirow}
\usepackage{float}
\usepackage{graphicx}
\journal{arXiv.org}
\begin{document}
\begin{frontmatter}

\title{An Efficient Method For Solving Highly Anisotropic Elliptic Equations}

\author[Phillips]{Edward Santilli\corref{cor1}}
\ead{santilli@physics.unc.edu}
\address[Phillips]{Dept. of Physics, 214 Phillips Hall, UNC, Chapel Hill, NC 27599}
\cortext[cor1]{Corresponding author}

\author[Venable]{Alberto Scotti}
\ead{ascotti@unc.edu}
\address[Venable]{Dept. of Marine Sciences, 3117H Venable Hall, UNC, Chapel Hill, NC 27599}

\begin{abstract}
  Solving elliptic PDEs in more than one dimension can be a computationally
  expensive task. For some applications characterised by a high degree of
  anisotropy in the coefficients of the elliptic operator, such that the term
  with the highest derivative in one direction is much larger than the terms in
  the remaining directions, the discretized elliptic operator often has a very
  large condition number -- taking the solution even further out of reach using
  traditional methods. This paper will demonstrate a solution method for such
  ill-behaved problems. The high condition number of the $D$-dimensional
  discretized elliptic operator will be exploited to split the problem into a
  series of well-behaved one and $\left( D-1 \right)$-dimensional elliptic
  problems. This solution technique can be used alone on sufficiently coarse
  grids, or in conjunction with standard iterative methods, such as Conjugate
  Gradient, to substantially reduce the number of iterations needed to solve the
  problem to a specified accuracy. The solution is formulated analytically for a
  generic anisotropic problem using arbitrary coordinates, hopefully bringing
  this method into the scope of a wide variety of applications.
\end{abstract}

\begin{keyword}
  \PACS 02.60.Lj \sep 47.11.St
  \MSC 35J57 \sep 65F10 \sep 65F08 \sep 65N22
\end{keyword}
\end{frontmatter}

\newpage
\tableofcontents
\newpage
\section{Introduction}

The solution of Poisson problems is often the single most expensive step in the
numerical solution of partial differential equations (PDEs). For example, when
solving the Navier-Stokes or Euler equations, the Poisson problem arises from
the incompressibility condition \cite{Minion, cyclic_reduction}. The particular
solution strategy depends of course on a combination of factors, including the
specific choice of the discretization and the type of boundary conditions. In
simple geometries, very efficient schemes can be devised to reduce the effective
dimensionality of the problem, such as using FFTs or cyclic reduction to
partially or completely diagonalize the operator \cite{cyclic_reduction}.
For more complex geometries or boundary conditions, the available choices to
solve the discretized problem usually involves direct inversion for small size
problems, while Krylov based iterative methods such as Conjugate Gradient or
multigrid methods are used when the matrix problem is too large to be inverted
exactly \cite{cyclic_reduction,Demmel,suntans}.

In this paper, we are interested in Poisson problems characterized by a high
level of anisotropy (to be precisely defined later). The source of anisotropy
can be due to the highly flattened domains over which the solution is sought, as
is the case for atmospheric, oceanic \cite{cyclic_reduction} and some
astrophysical problems \cite[][p. 77]{BinneyTremaine}. However, the source of
anisotropy could have a physical base, e.g. Non-Fickian diffusion problems where
the flux is related to the gradient by an anisotropic tensor \cite{nonfickian}.
Recently, research has been conducted to develop compound materials that could
serve as a cloaking device. \cite{cloak, magcloak} These \textit{metamaterials}
are designed to have specific anisotropic acoustic and electromagnetic
properties that divert pressure and light waves around a region of space
unscathed.

\begin{figure}[h!]
  \centering
  \includegraphics[width=\textwidth]{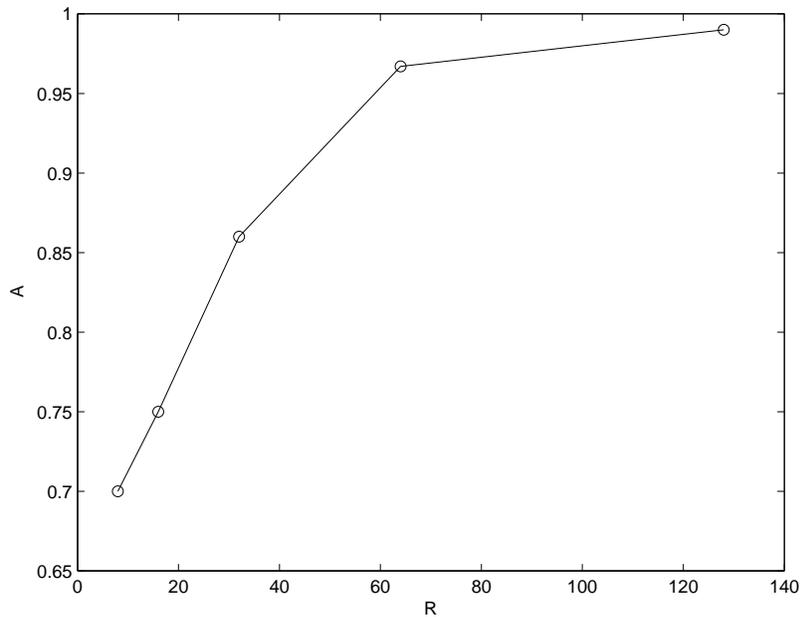}
  \caption{Attenuation factor as a function of domain aspect ratio for a Poisson
  problem solved used a standard multigrid scheme.}
  \label{fig_error}
\end{figure}

Anisotropy results in a spreading of the spectrum of the discretized operator,
with severe consequences on the convergence rate. We illustrate this point with
a simple Poisson problem
\begin{equation*}
  \nabla^2 \phi = \rho.
\end{equation*}
The r.h.s. and boundary conditions are chosen randomly (but compatible, see
below). The Laplacian operator is discretized with a standard 7-point,
second-order stencil, the domain is rectangular with dimensions $L \times L
\times H$, the domain's aspect ratio is measured by $R = L/H$, and the
discretization is chosen as $\left( \Delta x, \Delta y, \Delta z \right) =
\left( H/2, H/2, H/16 \right)$. For illustration
purposes, the problem is solved using a 4-level multigrid scheme which employs
line relaxation in the vertical (stiff) direction, as used by Armenio and Roman
\cite{ArmenioR09} to do a LES of a shallow coastal area. Figure \ref{fig_error}
shows the attenuation factor $A = \|E_{n + 1} \|/\|E_n \|$ as a function
of the aspect ratio, where $E_n$ is the residual error after $n$ iterations. For
moderate aspect-ratio domains, the convergence is satisfactory, but as $R$
increases, we rapidly approach a point where the method becomes, for practical
purposes, useless. Similar results (see below) hold for Krylov based methods.

In this paper, we describe how a formal series solution of a Poisson problem
derived by Scotti and Mitran \cite{Scotti}, herein referred to as SM, can be
used to significantly speed up the convergence of traditional iterative schemes.
SM introduced the concept of grid lepticity, $\lambda$, to describe the degree
of anisotropy of a discretized domain and then sought a solution to the Poisson
problem written as a power series in $\lambda^{- 1}$ -- the leptic expansion. An
apparent limitation of the leptic expansion is that it is very efficient only
for lepticity larger than a critical value of order $1$. SM were led to
introduce the leptic expansion in order to provide the right amount of
dispersion needed to balance nonlinear steepening of internal waves propagating
in a shallow stratified ocean. For this limited purpose, SM showed that at most
only three terms in the expansion are needed, and thus the lack of overall convergence was not a serious limitation. Here, we develop the method for the
purpose of efficiently calculating solutions of a discretized Poisson problem.
In our approach, the lepticity, which in SM's original formulation was related
to the aspect ratio of the domain, becomes now a generic measure of anisotropy.
The main result of this paper is that for subcritical values of the lepticity,
the leptic expansion can still be extremely valuable to dramatically increase
the convergence rate of standard iterative schemes, as the numerical
demonstrations of the method will show. The examples are coded using Matlab and
Chombo's BoxTools{\footnote{Chombo has been developed and is being distributed
by the Applied Numerical Algorithms Group of Lawrence Berkeley National Lab.
Website: https://seesar.lbl.gov/ANAG/chombo/}} library with standard second
order discretization techniques. 

What makes the leptic expansion particularly attractive is that it can be
parallelized in a very straightforward way, as long as the decomposition of the
domain does not split along the stiff (vertical, in our examples) direction. For
comparison, the parallel implementation of the Incomplete Cholesky Decomposition
of a sparse matrix, which is used as a preconditioner for Conjugate Gradient
schemes and yields very good convergence rates even at high levels of anisotropy
\cite{BruasetT94}, is a highly non-trivial task \cite{TeranishiR07}.

Finally, it must be noted that the idea behind the leptic expansion can be
traced as far back as the work of Bousinnesq on waves \cite{Boussinesq}. What we
have done here is to formulate it in a way suitable for numerical calculations.

The rest of the paper is organized as follows: a discretization- and
coordinate-independent version of the leptic expansion is introduced in Section
\ref{Derivation}. The reader who is not interested in the details may skip
directly to Section \ref{Expansion_Summary}, where a summary of the scheme is
provided. Section \ref{ref_convergence} presents convergence estimates of the
leptic expansion using Fourier analysis techniques \cite{ChanE89}. This is where
the leptic method's potential to generate initial guesses for conventional
iterative schemes emerges. In Section \ref{sec_dem}, we consider some examples
to illustrate how the leptic expansion can be used with conventional iterative
schemes to create very efficient solvers. A final section summarizes the main
results. 

\subsection{Notation}

As an aid to the following discussion, we define the relevant notation here.
\begin{itemize}
  \item Horizontal coordinate directions: $x^1 = x, x^2 = y$. Vertical (stiff) coordinate direction: $x^3 = z$.
  
  \item $H =$ vertical domain extent. $L =$ horizontal domain extent.
  
  \item $\phi =$ full solution. $\phi_p =$ solution of $p^{\text{th}}$-order
  equation.
  
  \item $\phi^v$ and $\phi^h =$ solution of vertical and horizontal equations
  (explained in section \ref{Derivation}).
  
  \item Summation indices\footnote{I will use summation notation. Repeated
  indices imply a sum unless explicitly stated otherwise.} (summed from $1$ to
  $3$): $i,j$.
  
  \item Horizontal summation indices (summed from $1$ to $2$): $m,n$.
  
  \item $u^{\star i} = \left( u^{\star},v^{\star},w^{\star} \right)^i$, the flux
  field that will be used as Neumann boundary conditions.
  
  \item $\rho =$ the source term of the elliptic PDE.
  
  \item $\sigma^{ij} =$ a symmetric, contravariant tensor field.
  
  \item $\mathcal{V}$ is the 3-dimensional domain and $d\mathcal{V} = dx^1 dx^2
  dx^3$.
  
  \item $\mathcal{A}_i$ is the boundary of $\mathcal{V}$ in the
  $i^{\text{th}}$-direction.
  
  \item $d\mathcal{A}_i = \left( dx^2 dx^3, dx^3 dx^1, dx^1 dx^2 \right)^T_i$,
  the area element at the boundary of $\mathcal{V}$.
  
  \item $\mathcal{S}$ is the 2-dimensional horizontal domain with local
  coordinates $\left( x^1,x^2 \right)$ and $d\mathcal{S}= dx^1 dx^2$.
  
  \item $dl_m = \left( dx^1,dx^2 \right)^T_m$, the line element around the
  boundary of $\mathcal{S}$.
  
  \item $\bar{A} \left( x,y \right) = \frac{1}{H} \int_{z_-}^{z_+} A \left(
  x,y,z \right) dz$, the vertical average of $A$.
  
  \item The leptic ratio is defined as $\lambda = \Delta x/H$, where
  $\Delta x$ is the horizontal mesh spacing.
\end{itemize}

\section{Derivation}\label{Derivation}

The problem we wish to solve is an anisotropic elliptic PDE with Neumann
boundary conditions of the type that often arises when solving the
incompressible Navier-Stokes equations \cite{Minion, cyclic_reduction}. More
precisely, we wish to solve the following equation
\begin{eqnarray}
  \label{Poisson}
  \partial_i \sigma^{ij} \partial_j \phi
  & = & \rho \text{ in } \mathcal{V}
  \\
  \sigma^{ij} \partial_j \phi
  & = & u^{\star i} \text{ on } \partial \mathcal{V}, \nonumber
\end{eqnarray}
where $\sigma^{ij}$ is a positive-definite, symmetric tensor field. The only
restriction on $\rho$ and $u^{\star i}$ is that they must be compatible with one
another. That is, if we integrate eq. (\ref{Poisson}) over $\mathcal{V}$ and
apply Stokes' theorem, we obtain an identity that must be obeyed by the sources,
\begin{equation}
  \label{compat}
  \int_{\mathcal{V}} \rho d\mathcal{V}
  = \oint_{\partial \mathcal{V}} u^{\star i} d\mathcal{A}_i.
\end{equation}

In general, the domain can have any number of dimensions higher than 1, but
without loss of generality we will restrict ourselves to the 3-dimensional
case. We will also assume there is a small parameter, $\varepsilon$, that we can
use to identify terms of the field and operator in a formal perturbation
expansion of the form
\begin{eqnarray}
  \label{expansion}
  \phi
  & = & \phi_0
  + \varepsilon \phi_1
  + \varepsilon^2 \phi_2
  + \varepsilon^3 \phi_3
  + \ldots
  \\
  \partial_i \sigma^{ij} \partial_j
  & = & \partial_3 \sigma^{33} \partial_3
  + \varepsilon \left( \partial_m \sigma^{mn} \partial_n
  + \partial_m \sigma^{m3} \partial_3
  + \partial_3 \sigma^{3n} \partial_n \right). \nonumber
\end{eqnarray}
In this section, we will derive a method to solve eqs. (\ref{Poisson}) using the
expansion (\ref{expansion}).

\subsection{The desired form of the expansion}

We begin by plugging expansion (\ref{expansion}) into the first of eqs.
(\ref{Poisson}) and equating powers of $\varepsilon$.
\begin{eqnarray}
  \label{O0raw}
  \partial_3 \sigma^{33} \partial_3 \phi_0 & = & \rho
  \\
  \label{O1raw}
  \partial_3 \sigma^{33} \partial_3 \phi_1
  + \left( \partial_m \sigma^{m3} \partial_3
  + \partial_3 \sigma^{3n} \partial_n
  + \partial_m \sigma^{mn} \partial_n \right) \phi_0
  & = & 0
  \\
  \label{O2raw}
  \partial_3 \sigma^{33} \partial_3 \phi_2
  + \left( \partial_m \sigma^{m3} \partial_3
  + \partial_3 \sigma^{3 n} \partial_n
  + \partial_m \sigma^{mn} \partial_n \right) \phi_1
  & = & 0
  \\
  \text{etc} \ldots
  &  &  \nonumber
\end{eqnarray}
The first thing to notice is that eq. (\ref{O0raw}) with Neumann boundary
conditions can only determine $\phi_0$ up to an additive function of $x$ and
$y$. We will call the solution of eq. (\ref{O0raw}) $\phi_0^v(x,y,z)$ and the
still undetermined function $\phi_0^h(x,y)$. We might as well preemptivley write
the fields at every order as
\begin{equation*}
  \phi_n(x,y,z) = \phi_n^v(x,y,z) + \phi_n^h(x,y)
\end{equation*}
so that equations (\ref{O0raw})-(\ref{O2raw}) read
\begin{eqnarray}
  \label{O0}
  \partial_3 \sigma^{33} \partial_3 \phi_0^v & = & \rho
  \\
  \label{O1}
  \partial_3 \sigma^{33} \partial_3 \phi_1^v
  + \partial_m \sigma^{m3} \partial_3 \phi_0^v
  + \left( \partial_3 \sigma^{3n} \partial_n
  + \partial_m \sigma^{mn} \partial_n \right) \phi_0
  & = & 0
  \\
  \label{O2}
  \partial_3 \sigma^{33} \partial_3 \phi_2^v
  + \partial_m \sigma^{m3} \partial_3 \phi_1^v
  + \left( \partial_3 \sigma^{3n} \partial_n
  + \partial_m \sigma^{mn} \partial_n \right) \phi_1
  & = & 0
  \\
  \text{etc} \ldots
  &  &  \nonumber
\end{eqnarray}

In order to solve this set of equations, we must define our boundary conditions
at each order. At $\mathcal{O}(1)$, we will set
\begin{eqnarray*}
  \left. \sigma^{33} \partial_3 \phi_0^v \right|_{z_+}
  & = & \left. w^{\star} \right|_{z_+} - \tilde{w}_0
  \\
  \left. \sigma^{33} \partial_3 \phi_0^v \right|_{z_-}
  & = & \left. w^{\star} \right|_{z_-},
\end{eqnarray*}
where $z_+$ and $z_-$ denote the evaluation at the upper and lower boundaries,
respectively. The excess function, $\tilde{w}_0$, is defined at each $x$ and $y$
to make eq. (\ref{O0}) consistent with its boundary conditions. By vertically
integrating eq. (\ref{O0}), we see that
\begin{eqnarray*}
  \int_{z_-}^{z_+} \rho dz
  & = & \left. \sigma^{33} \partial_3 \phi_0^v \right|_{z_-}^{z_+}
  \\
  & = & \left. w^{\star} \right|_{z_-}^{z_+} - \tilde{w}_0,
\end{eqnarray*}
that is,
\begin{equation*}
  \tilde{w}_0 = \left. w^{\star} \right|_{z_-}^{z_+} - \int_{z_-}^{z_+} \rho dz.
\end{equation*}
This completely determines $\phi_0^v$ along vertical lines at each $x$ and $y$.
Notice that we have not yet chosen the gradients of $\phi_0^v$ normal to the
horizontal boundaries. We will save this freedom for later. Right now, we need
to look at the $\mathcal{O}(\varepsilon)$ equations to get $\phi_0^h$.

For the moment, let us think of eq. (\ref{O1}) as an equation for $\phi_1^v$.
We again need to specify vertical boundary conditions. We will define
\begin{eqnarray}
  \label{phi1vBCs}
  \left. \sigma^{33} \partial_3 \phi_1^v \right|_{z_+}
  & = & \tilde{w}_0 - \left. \sigma^{3n} \partial_n \phi_0 \right|_{z_+}
  \\
  \left. \sigma^{33} \partial_3 \phi_1^v \right|_{z_-}
  & = & - \left. \sigma^{3n} \partial_n \phi_0 \right|_{z_-}. \nonumber
\end{eqnarray}
Defining a new excess function is unnecessary because it can just be absorbed
into the still undetermined function $\phi_0^h$. As before, we vertically
integrate eq. (\ref{O1}).
\begin{eqnarray*}
  0
  & = & \left[ \sigma^{33} \partial_3 \phi_1^v
  + \sigma^{3n} \partial_n \phi_0 \right]_{z_-}^{z_+}
  + \int_{z_-}^{z_+} \left( \partial_m \sigma^{m3} \partial_3 \phi_0^v
  + \partial_m \sigma^{mn} \partial_n \phi_0 \right) dz
  \\
  & = & \tilde{w}_0
  + \int_{z_-}^{z_+} \left( \partial_m \sigma^{m3} \partial_3 \phi_0^v
  + \partial_m \sigma^{mn} \partial_n \phi_0 \right) dz
  \\
  & = & \tilde{w}_0
  + \int_{z_-}^{z_+} \left( \partial_m \sigma^{m3} \partial_3 \phi_0^v
  + \partial_m \sigma^{mn} \partial_n \phi_0^v \right) dz
  + \int_{z_-}^{z_+} \partial_m \sigma^{mn} \partial_n \phi_0^h dz
  \\
  & = & \tilde{w}_0
  + \int_{z_-}^{z_+} \partial_m \sigma^{mj} \partial_j \phi_0^v dz
  + \int_{z_-}^{z_+} \partial_m \sigma^{mn} \partial_n \phi_0^h dz
\end{eqnarray*}
If we divide by $H$, the vertical integrals become vertical averages, which will
be denoted with overbars. When taking these averages, remember that while
$\phi^h$ was defined to be independent of $z$, no such assumption was made for
$\sigma^{ij}$. This leaves us with an equation for $\phi_0^h$,
\begin{equation}
  \label{phi0h}
  \partial_m \overline{\sigma^{mn}} \partial_n \phi_0^h
  = - \frac{\tilde{w}_0}{H}
  - \partial_m \overline{\sigma^{mj} \partial_j \phi_0^v}.
\end{equation}
If $\phi_0^h$ is chosen to be any solution to this equation, then eq. (\ref{O1})
for $\phi_1^v$ together with the boundary conditions (\ref{phi1vBCs}) will be
consistent. Now, we define boundary conditions for eq. (\ref{phi0h}) to be
\begin{equation}
  \label{phi0hBCs}
  \left. \overline{\sigma^{mn}} \partial_n \phi_0^h \right|
  _{\vec{x} \in \partial\mathcal{S}}
  = \left. \overline{u^{\star m}} \right|_{\vec{x} \in \partial \mathcal{S}}.
\end{equation}
This choice of boundary condition will be made consistent with equation
(\ref{phi0h}) in the following steps. First, we integrate eq. (\ref{phi0h}) over
$\mathcal{S}$ and reorganize the result.
\begin{eqnarray*}
  - \frac{1}{H} \int_{\mathcal{S}} \tilde{w}_0 d\mathcal{S} 
  - \oint_{\partial \mathcal{S}} \overline{\sigma^{mj} \partial_j \phi_0^v} dl_m
  & = & - \int_{\mathcal{S}} \frac{\tilde{w}_0}{H} d\mathcal{S}
  - \int_{\mathcal{S}} \partial_m
  \overline{\sigma^{mj} \partial_j \phi_0^v} d\mathcal{S}
  \\
  & = & \int_{\mathcal{S}} \partial_m \overline{\sigma^{mn}} 
  \partial_n \phi_0^h d\mathcal{S}
  \\
  & = & \oint_{\partial \mathcal{S}} \overline{\sigma^{mn}} 
  \partial_n \phi_0^h dl_m
  \\
  & = & \oint_{\partial \mathcal{S}} \overline{u^{\star m}} dl_m
\end{eqnarray*}
Next, we exploit the remaining freedoms in the boundary conditions of $\phi_0^v$
by choosing $\sigma^{mj} \partial_j \phi_0^v = u^{\star m} -
\overline{u^{\star m}}$ at the horizontal boundaries of $\mathcal{V}$. This,
together with the definition of $\tilde{w}_0$, gives us
\begin{equation*}
  - \frac{1}{H} \int_{\mathcal{S}} \left( \left. w^{\star} \right|_{z_-}^{z_+}
  - \int_{z_-}^{z_+} \rho dz \right) d\mathcal{S}
  - \oint_{\partial S} \overline{\left( u^{\star m}
  - \overline{u^{\star m}} \right)} dl_m
  = \oint_{\partial \mathcal{S}} \overline{u^{\star m}} dl_m.
\end{equation*}
Noting that the third and fourth terms cancel, this simplifies to become
\begin{equation*}
  H \oint_{\partial \mathcal{S}} \overline{u^{\star m}} dl_m
  + \int_{\mathcal{S}} \left. w^{\star} \right|_{z_-}^{z_+} d\mathcal{S}
  = \int_{\mathcal{V}} \rho d\mathcal{V}.
\end{equation*}
Now, if this equation holds, then the boundary conditions (\ref{phi0hBCs}) will
be consistent with the horizontal equation (\ref{phi0h}). From equation
(\ref{compat}), we see that this is indeed the case. Therefore, equations
(\ref{phi0h}) and (\ref{phi0hBCs}) completely determine $\phi_0^h$, and in turn,
$\phi_0$. Having $\phi_0$ at our disposal, we can now tidy up eq. (\ref{O1}) a
bit.
\begin{eqnarray*}
  \partial_3 \sigma^{33} \partial_3 \phi_1^v
  & = & - \partial_m \sigma^{m3} \partial_3 \phi_0^v 
  - \left( \partial_m \sigma^{mn} \partial_n 
  + \partial_3 \sigma^{3n} \partial_n \right) \phi_0
  \\
  & = & \left( \rho - \partial_3 \sigma^{33} \partial_3 \phi_0 \right) 
  - \partial_m \sigma^{m3} \partial_3 \phi_0 
  - \left( \partial_m \sigma^{mn} \partial_n 
  + \partial_3 \sigma^{3n} \partial_n \right) \phi_0
  \\
  & = & \rho - \partial_i \sigma^{ij} \partial_j \phi_0
\end{eqnarray*}
The first two terms in parentheses in the second line are zero via eq.
(\ref{O0}). Since this equation along with boudary conditions (\ref{phi1vBCs})
are consistent, this completely determines $\phi_1^v$ at each $x$ and $y$. There
is, as before, another freedom yet to be chosen -- the gradients of $\phi_1^v$
normal to the horizontal boundaries. Again, we will choose them later.

We continue in the same manner, obtaining an equation for $\phi_2^v$ whose
Neumann compatibility condition is the equation for $\phi_1^h$. At this point,
we might as well just derive the equations for the general order fields
$\phi_p^v$ and $\phi_{p - 1}^h$ where $p \geq 2$. We start with
\begin{equation}
  \label{Op}
  \partial_3 \sigma^{33} \partial_3 \phi_p^v
  + \partial_m \sigma^{m3} \partial_3 \phi_{p-1}^v
  + \left( \partial_m \sigma^{mn} \partial_n
  + \partial_3 \sigma^{3n} \partial_n \right) \phi_{p-1}
  = 0
\end{equation}
and the vertical boundary conditions for $\phi_p^v$
\begin{equation*}
  \left. \sigma^{33} \partial_3 \phi_p^v \right|_{z_{\pm}} 
  = - \left. \sigma^{3n} \partial_n \phi_{p-1} \right|_{z_{\pm}}.
\end{equation*}
Vertically averaging eq. (\ref{Op}) and rearranging a bit gives us
\begin{equation*}
  \frac{1}{H} \left[ \sigma^{33} \partial_3 \phi_p^v 
  + \sigma^{3n} \partial_n \phi_{p-1} \right]_{z_-}^{z_+}
  + \overline{\partial_m \sigma^{mj} \partial_j \phi_{p-1}^v 
  + \partial_m \sigma^{mn} \partial_n \phi^h_{p - 1}} 
  = 0,
\end{equation*}
which upon applying boundary conditions and rearranging some more gives us the 
horizontal equation
\begin{equation}
  \label{phiph}
  \partial_m \overline{\sigma^{mn}} \partial_n \phi^h_{p-1} =
  - \partial_m \overline{\sigma^{mj} \partial_j \phi_{p-1}^v}.
\end{equation}

This equation will be compatible with homogeneous boundary conditions if we 
chose the gradients of $\phi^v_{p}$ at the horizontal boundaries wisely. By 
integrating eq. (\ref{phiph}) over $\mathcal{S}$, we find
\begin{eqnarray*}
  0 & = & \oint_{\partial\mathcal{S}} \overline{\sigma^{mn}}
  \partial_{n} \phi^h_{p-1} dl_{m}
  \\
  & = & \int_{\mathcal{S}} \partial_{m} \overline{\sigma^{mn}}
  \partial_{n} \phi^h_{p-1} d\mathcal{S}
  \\
  & = & -\int_{\mathcal{S}} \partial_{m} \overline{\sigma^{mj}
  \partial_{j} \phi^v_{p-1}} d\mathcal{S}
  \\
  & = & -\oint_{\partial\mathcal{S}} \overline{\sigma^{mj}
  \partial_{j} \phi^v_{p-1}} dl_{m}.
\end{eqnarray*}
It is tempting now to let the gradients of $\phi^v_{p-1}$ for $p\geq 2$ be 
identically zero, but a more appropriate choice is
\begin{equation}
  \label{vertical_gradients}
  \sigma^{mj} \partial_{j} \phi_{p-1}^v
  = \left(\overline{\sigma^{mn}} - \sigma^{mn}\right) \partial_n \phi^h_{p-2},
\end{equation}
which is equivalent to
\begin{equation*}
  \sigma^{mj} \partial_{j} \phi_p^v =
  \left\{ \begin{array}{ll}
    \overline{u^{\star m}} - \sigma^{mn} \partial_n \phi^h_{0}, & p = 1
    \\
    - \sigma^{mn} \partial_n \phi^h_{p-1}, & p \geq 2.
  \end{array} \right.
\end{equation*}
This choice will average to zero, satisfying the above integral, as well as
prevent inconsistencies later.

Finally, we can clean up eq. (\ref{Op}) to get an equation for $\phi_p^v$.
\begin{equation*}
  \partial_3 \sigma^{33} \partial_3 \phi_p^v
  = \rho - \partial_i \sigma^{ij} \partial_j
  \left( \phi_0 + \phi_1 + \ldots + \phi_{p-1} \right)
\end{equation*}
This completes the derivation of the needed equations. It may seem like some
of the boundary conditions were chosen only to be compatible with their
corresponding differential equation, when in fact we chose them carefully so
that the sum of their contributions is $u^{\star i}$ for each direction, $i$.
At the horizontal boundaries,
\begin{align*}
  \sigma^{mj} \partial_j \phi
  &= \left( \sigma^{mj} \partial_j \phi_0^v \right)
  + \left( \sigma^{mn} \partial_n \phi_0^h
  + \sigma^{mj} \partial_j \phi_1^v \right)
  + \ldots
  + \left( \sigma^{mn} \partial_n \phi_{p-1}^h
  + \sigma^{mj} \partial_j \phi_p^v \right)
  + \ldots
  \\
  &= \left( u^{\star m} - \overline{u^{\star m}} \right)
  + \left( \overline{\sigma^{mn}} \partial_{n} \phi^h_0 \right)
  + \ldots
  + \left( \overline{\sigma^{mn}} \partial_{n} \phi^h_{p-1} \right)
  + \ldots
  \\
  &= \left( u^{\star m} - \overline{u^{\star m}} \right)
  + \left( \overline{u^{\star m}} \right)
  + \ldots
  + \left( 0 \right)
  + \ldots
  \\
  &= u^{\star m},
\end{align*}
at the upper vertical boundary,
\begin{align*}
  \left. \sigma^{3j} \partial_j \phi \right|_{z_+}
  &= \left( \sigma^{33} \partial_3 \phi_0^v \right)
  + \left( \sigma^{3n} \partial_n \phi_0
  + \sigma^{33} \partial_3 \phi_1^v \right)
  + \ldots
  + \left( \sigma^{3n} \partial_n \phi_{p-1}
  + \sigma^{33} \partial_3 \phi_p^v \right)
  + \ldots
  \\
  &= \left( w^{\star} - \tilde{w}_0 \right)
  + \left( \tilde{w}_0 \right)
  + \ldots
  + \left( 0 \right)
  + \ldots
  \\
  &= w^{\star},
\end{align*}
and at the lower vertical boundary,
\begin{align*}
  \left. \sigma^{3j} \partial_j \phi \right|_{z_-}
  &= \left( \sigma^{33} \partial_3 \phi_0^v \right)
  + \left( \sigma^{3n} \partial_n \phi_0
  + \sigma^{33} \partial_3 \phi_1^v \right)
  + \ldots
  + \left( \sigma^{3n} \partial_n \phi_{p-1}
  + \sigma^{33} \partial_3 \phi_p^v \right)
  + \ldots
  \\
  &= \left( w^{\star} \right)
  + \left( 0 \right)
  + \ldots
  + \left( 0 \right)
  + \ldots
  \\
  &= w^{\star}.
\end{align*}

\subsection{Summary of the expansion}\label{Expansion_Summary}

Tables (\ref{tbl_general}) and (\ref{tbl_Cartesian}) provide a concise overview
of the steps involved in generating the $n^{th}$-order of the expansion in
generalized and Cartesian coordinates, respectively. The problem is formulated
recursively. The left hand side is the same at each step. However, it must be
noted that the solution of the $\left( D-1 \right)$-dimensional Poisson problem
can be postponed or in some cases eliminated depending on the magnitude of the
correction required. This will be discussed in section \ref{sec_elim_horiz}. For
a very simple example problem solved using this expansion, see section
\ref{ref_convergence}.
\begin{table}[H]
\centering
\renewcommand{\arraystretch}{1.75}
\resizebox{\textwidth}{!}{
\begin{tabular}{|l|l|l|}
  \hline
  \multirow{3}{*}{$\mathcal{O} \left( 1 \right)$}
  & \multirow{2}{*}{$\partial_3 \sigma^{33} \partial_3 \phi^v_0 = \rho$}
  & $\left. \sigma^{33} \partial_3 \phi^v_0 \right|_{z_\pm} =
  \left\{ \begin{array}{ll}
    \left. w^{\star} \right|_{z_+} - \tilde{w}_0, & z = z_+
    \\
    \left. w^{\star} \right|_{z_-}, & z = z_-
  \end{array} \right.$ \\
  & & $\left. \sigma^{mj}\partial_j \phi^v_0 \right|_{\vec{x} \in \mathcal{A}_m} =
  \left[ u^{\star m} - \overline{u^{\star m}} \right]_{\vec{x} \in \mathcal{A}_m}$ \\
  \cline{2-3}
  & $\partial_m \overline{\sigma^{mn}} \partial_n \phi^h_0 = -\frac{\tilde{w}_0}{H} - \partial_m \overline{\sigma^{mj} \partial_j \phi^v_0}$
  & $\left. \overline{\sigma^{mn}} \partial_n \phi^h_0 \right|_{\vec{x} \in \mathcal{\partial S}} = \left. \overline{u^{\star m}} \right|_{\vec{x} \in \mathcal{\partial S}}$\\
  \hline
  \multirow{3}{*}{$\mathcal{O} \left( \varepsilon \right)$}
  & \multirow{2}{*}{$\partial_3 \sigma^{33} \partial_3 \phi^v_1 = \rho - \partial_i \sigma^{ij} \partial_j \phi_0$}
  & $\left. \sigma^{33} \partial_3 \phi^v_1 \right|_{z_\pm} =
  \left\{ \begin{array}{ll}
    \tilde{w}_0 - \left. \sigma^{3n} \partial_n \phi_0 \right|_{z_+}, & z = z_+
    \\
    \left. -\sigma^{3n} \partial_n \phi_0 \right|_{z_-}, & z = z_-
  \end{array} \right.$ \\
  & & $\left. \sigma^{mj}\partial_j \phi^v_1 \right|_{\vec{x} \in \mathcal{A}_m} =
  \left[ \overline{u^{\star m}} - \sigma^{mn} \partial_n \phi^h_0 \right]_{\vec{x} \in \mathcal{A}_m}$ \\
  \cline{2-3}
  & $\partial_m \overline{\sigma^{mn}} \partial_n \phi^h_1 = -\partial_m \overline{\sigma^{mj} \partial_j \phi^v_1}$
  & $\left. \overline{\sigma^{mn}} \partial_n \phi^h_1 \right|_{\vec{x} \in \mathcal{\partial S}} = 0$\\
  \hline
  \multirow{3}{*}{$\mathcal{O} \left( \varepsilon^p \right)$}
  & \multirow{2}{*}{$\partial_3 \sigma^{33} \partial_3 \phi^v_p = \rho - \partial_i \sigma^{ij} \partial_j \left( \displaystyle\sum_{r=0}^{p-1}\phi_r \right)$}
  & $\left. \sigma^{33} \partial_3 \phi^v_p \right|_{z_\pm} = \left. -\sigma^{3n} \partial_n \phi_{p-1} \right|_{z_\pm} $\\
  & & $\left. \sigma^{mj}\partial_j \phi^v_p \right|_{\vec{x} \in \mathcal{A}_m} =
  \left. - \sigma^{mn} \partial_n \phi^h_{p-1} \right|_{\vec{x} \in \mathcal{A}_m}$ \\
  \cline{2-3}
  & $\partial_m \overline{\sigma^{mn}} \partial_n \phi^h_p = -\partial_m \overline{\sigma^{mj} \partial_j \phi^v_p}$
  & $\left. \overline{\sigma^{mn}} \partial_n \phi^h_p \right|_{\vec{x} \in \mathcal{\partial S}} = 0$\\
  \hline
\end{tabular} }
\caption[The general form of the expansion]{The general form of the expansion. The indices $i,j$ extend over all directions and the indices $m,n$ do not include the vertical (thin) direction. The excess function is defined as $\tilde{w}_0 = \left. w^{\star} \right|_{z_-}^{z_+} - \int_{z_-}^{z_+} \rho dz$.}
\label{tbl_general}
\end{table}

\begin{table}[H]
\centering
\renewcommand{\arraystretch}{1.75}
\begin{tabular}{|l|l|l|}
  \hline
  \multirow{3}{*}{$\mathcal{O} \left( 1 \right)$}
  & \multirow{2}{*}{$\partial_z^2 \phi^v_0 = \rho$}
  & $\left. \partial_z \phi^v_0 \right|_{z_\pm} =
  \left\{ \begin{array}{ll}
    \left. w^{\star} \right|_{z_+} - \tilde{w}_0, & z = z_+
    \\
    \left. w^{\star} \right|_{z_-}, & z = z_-
  \end{array} \right.$ \\
  & & $\left. \partial_m \phi^v_0 \right|_{\vec{x} \in \mathcal{A}_m} = \left[ u^{\star m} - \overline{u^{\star m}} \right]_{\vec{x} \in \mathcal{A}_m}$ \\
  \cline{2-3}
  & $\nabla_h^2 \phi^h_0 = -\frac{\tilde{w}_0}{H}$
  & $\left. \partial_m \phi^h_0 \right|_{\vec{x} \in \mathcal{\partial S}} = \left. \overline{u^{\star m}} \right|_{\vec{x} \in \mathcal{\partial S}}$\\
  \hline
  \multirow{3}{*}{$\mathcal{O} \left( \varepsilon \right)$}
  & \multirow{2}{*}{$\partial_z^2 \phi^v_1 = \rho - \nabla^2 \phi_0$}
  & $\left. \partial_z \phi^v_1 \right|_{z_\pm} =
  \left\{ \begin{array}{ll}
    \tilde{w}_0, & z = z_+
    \\
    0, & z = z_-
  \end{array} \right.$ \\
  & & $\left. \partial_m \phi^v_1 \right|_{\vec{x} \in \mathcal{A}_m} = 0$ \\
  \cline{2-3}
  & 
  \multicolumn{2}{l|}{No horizontal equation. $\phi^h_1 = 0$} \\
  \hline
  \multirow{3}{*}{$\mathcal{O} \left( \varepsilon^p \right)$}
  & \multirow{2}{*}{$\partial_z^2 \phi^v_p = \rho - \nabla^2 \left( \displaystyle\sum_{r=0}^{p-1}\phi_r \right)$}
  & $\left. \partial_z \phi^v_p \right|_{z_\pm} = 0 $\\
  & & $\left. \partial_m \phi^v_p \right|_{\vec{x} \in \mathcal{A}_m} = 0$ \\
  \cline{2-3}
  & \multicolumn{2}{l|}{No horizontal equation. $\phi^h_p = 0$} \\
  \hline
\end{tabular}
\caption[The expansion in Cartesian coordinates]{The expansion in Cartesian coordinates. The indices $i,j$ extend over all directions and the indices $m,n$ do not include the vertical (thin) direction. The excess function is defined as $\tilde{w}_0 = \left. w^{\star} \right|_{z_-}^{z_+} - \int_{z_-}^{z_+} \rho dz$ and the horizontal Laplacian is $\nabla_h^2 = \partial_x^2 + \partial_y^2$.}
\label{tbl_Cartesian}
\end{table}

\subsection{Eliminating horizontal stages}\label{sec_elim_horiz}

Typically, solutions of the horizontal problems are more expensive than
solutions of the vertical problems. Sometimes, we can skip the horizontal
stages altogether if we know in advance that its solution will not contribute
to the overall convergence of the method. For example, suppose
\begin{equation*}
  \sigma^{ij} =
  \left( \begin{array}{ccc}
    A \left( x, y \right) & D \left( x, y \right) & 0\\
    D \left( x, y \right) & B \left( x, y \right) & 0\\
    0 & 0 & C \left( x, y, z \right)
   \end{array} \right)
\end{equation*}
where $A, B, C,$ and $D$ are arbitrary functions of the variables listed.
We see that
\begin{equation*}
  \partial_m \overline{\sigma^{mj} \partial_j \phi^v}
  = \partial_m \sigma^{mn} \partial_n \overline{\phi^v}
  + \partial_m \sigma^{m3} \overline{\partial_3 \phi^v},
\end{equation*}
but since $\sigma^{m3} = 0$ by assumption, the $\partial_m \sigma^{m3}
\overline{\partial_3 \phi^v}$ term is zero. Also, since each $\phi^v_p$ is the
solution of an ordinary differential equation with Neumann BCs, we can choose
solutions whose vertical averages are zero -- eliminating the $\partial_m
\sigma^{mn} \partial_n \overline{\phi^v}$ term as well. This means that
the r.h.s. of the horizontal equations become zero for all but the
$\mathcal{O}(1)$ stages. Since the boundary conditions are also zero for these
problems, the solutions, $\phi^h_{p \geq 1}$, must be identically zero.
Whenever $\sigma^{ij}$ has this form, $\phi^h_0$ is the only horizontal
function that needs to be found -- eliminating most of the computation time.

In practice, $\sigma^{ij}$ is often very close to the form shown above. In fact,
many useful coordinate systems such as the Cartesian, cylindrical, and spherical
systems are described by $\sigma^{ij} = \sqrt{g} g^{i j}$ which are
{\em exactly} of this form. It is helpful to consider this while iterating. If
we can find a value of $P$ such that all $\phi_{p \geq P}^h$ will not
significantly influence the overall convergence, or if we calculate that the
norm of the horizontal equation's r.h.s. is below some threshold, then we can
tell the leptic solver to stop performing horizontal solves in the interest of
computation time. Alternatively, suppose we are about to find $\phi^h_p$. We
could simply set $\phi^h_p$ to zero everywhere and then set up the next vertical
stage. If the vertical problem is consistent up to some prescribed tolerance,
then we never needed the true solution of the horizontal equation to begin with!
Otherwise, we can go back and solve the horizontal equation before moving on to
the next vertical stage. This is a very economical way of deciding which
horizontal solves are necessary.

\subsection{An interpretation of $\varepsilon$}

So far, we have derived a set of equations that produce a formal solution to the
original Poisson problem based on the assumption that the parameter
$\varepsilon$ in eqs. (\ref{expansion}) properly identifies terms of
fundamentally different sizes. The procedure does not refer to a specific
discretization of the equations, but heuristically depends on the existence of a
small parameter. The latter is typically derived from an anisotropy inherent in
the problem. This could be due to many causes, but to appeal to the interests of
the author's own research, we will focus on an anisotropy in the geometry of the
domain and numerical discretization. However, it is easy to generalize the
foregoing argument regardless of the actual source of anisotropy. Naively, the
aspect ratio of the domain could be used to define such a parameter. However,
the following example shows that it must also depend on the details of the
discretization.

Suppose we want to solve the isotropic, 2D Poisson equation in a rectangular
domain. We will choose a uniform discretization with $N_x \times N_z$
cell-centers and we will define the aspect ratio to be $\alpha = H/L$.
Upon switching to the dimensionless variables $\tilde{x} = x/L$ and
$\tilde{z} = z/H$, Poisson's equation transforms as follows
\begin{eqnarray*}
  \rho & = & \left( \frac{\partial^2}{\partial x^2} +
  \frac{\partial^2}{\partial z^2} \right) \phi
  \\
  & = & \frac{1}{H^2} \left( \alpha^2 L^2 \frac{\partial^2}{\partial x^2} +
  H^2 \frac{\partial^2}{\partial z^2} \right) \phi
  \\
  & = & \frac{1}{H^2} \left( \alpha^2 \frac{\partial^2}{\partial \tilde{x}^2}
  + \frac{\partial^2}{\partial \tilde{z}^2} \right) \phi
\end{eqnarray*}
We will rescale the field variable, $\phi$, so that it is dimensionless and of
$\mathcal{O} \left( 1 \right)$, which is always possible. Now, apart from an
overall scaling of $H^{-2}$, it is natural to identify the small
parameter $\varepsilon$ with the aspect ratio of the grid. However, a little
reflection shows that a more ``quantitative'' definition of $\varepsilon$
cannot ignore the discretization altogether. Indeed, once discretized, the term
involving $x$-derivatives is at most $\sim \alpha^2 N_x^2$ while the term
involving $z$-derivatives is at least $\sim 1$. This means that if $\alpha N_x
\ll 1$, then $\frac{\partial^2 \phi}{\partial x^2}$ will be fundamentally
smaller than $\frac{\partial^2 \phi}{\partial z^2}$, and so the ``smallness'' of
$\varepsilon$ depends on both the aspect ratio of the domain and on the details
of its discretization. It follows that, in the continuum limit, $\varepsilon$
cannot be {\em a priori} ever guaranteed to be small.

Scotti and Mitran quantified these results in a more general manner. In their
paper \cite{Scotti}, they define the grid's {\em leptic ratio}, $\lambda = \min
\left( \Delta x^m/H \right)$, where $\min \Delta x^m$ is the minimum grid
spacing in the directions other than the vertical. Note that the leptic ratio is
controlled by the overall apsect ratio of the domain and by the degree of
``coarseness'' of the discretization in the horizontal directions. It is shown
that if $\lambda > \mathcal{O}(1)$, then the computational grid is ``thin,'' and
the summarized equations of the previous section will indeed produce solutions
whose sum converges to the solution of (\ref{Poisson}). This result extends to
more exotic, N-dimensional geometries if we identify the symmetric tensor field,
$\sigma^{ij}$, with $\sqrt{g} g^{ij}$.

At first, it would seem that this method is of very limited practical utility,
being convergent only on rather coarse grids. Moreover, it it is at odds with
the very sensible idea that finer grids should lead to better results. However,
in this article we pursue the idea that even when the leptic ratio of the grid
is below critical, so that the expansion will not converge, the method can still
be used to accelerate the convergence of conventional methods. Looesely
speaking, the idea is that the r.h.s. can be partitioned between a component
that can be represented on a grid with $\lambda > 1$ and a remainder which needs
a finer grid with $\lambda < 1$. Restricted to the former component, the
expansion converges, while it diverges on the latter. On the contrary,
traditional method such as BiCGStab tend to converge fast on the latter, and
slow on the former. By judiciously blending both methods, we can achieve a
uniformly high level of convergence. Also, note that since the expansion is
formulated analytically, it can be implemented regardless of any particular
choice of discretization of the domain. In the following examples, we will use
a second-order scheme on a staggered grid, but the method is by no means
restricted to this type of discretization.

\section{Convergence estimates}\label{ref_convergence}

\subsection{Restricted case}

The heuristic arguments used earlier can be given an analytic justification
in the case of a simple rectangular geometry. Once again, we limit to two
dimensions, the extension to higher dimensional spaces being trivial. The
elliptic equation we wish to solve is
\begin{equation}
  \left( \partial_x^2 + \partial_z^2 \right) \phi \left( x,z \right)
  = \rho \left( x,z \right)
  \label{2DPoisson}
\end{equation}
with homogeneous Neumann boundary conditions. Without loss of generality, we can
set $\rho \left( x,z \right)$ equal to an arbitrary eigenfunction of the
operator, $\cos \left( kx \right) \cos \left( mz \right)$. The exact, analytic
solution of eq. (\ref{2DPoisson}) becomes simply
\begin{equation*}
  \phi \left( x,z \right)
  = - \frac{\cos \left( kx \right) \cos \left( mz \right)}{k^2 + m^2}.
\end{equation*}
Now, let's investigate how the leptic solver would have arrived at a solution.

First, we will write eq. (\ref{2DPoisson}) as
\begin{equation}
  \label{poisson_leptic}
  \left( \varepsilon \partial_x^2 + \partial_z^2 \right)
  \left( \phi_0 + \varepsilon \phi_1 + \ldots \right)
  = \cos \left( kx \right) \cos \left( mz \right),
\end{equation}
where the $\varepsilon$ is only used to identify small terms and will eventually
be set to $1$. Equating various powers of $\varepsilon$ gives us
\begin{eqnarray*}
  \partial_z^2 \phi_0 & = & \cos \left( kx \right) \cos \left( mz \right)\\
  \partial_z^2 \phi_1 & = & - \partial_x^2 \phi_0\\
  \partial_z^2 \phi_2 & = & - \partial_x^2 \phi_1\\
  & \text{etc} \ldots
\end{eqnarray*}
The solution to the $\mathcal{O} \left( 1 \right)$ equation is
\begin{equation*}
  \phi_0 \left( x,z \right)
  = \frac{\cos \left( kx \right) \cos \left( mz \right)}{-m^2}.
\end{equation*}
The constant of integration, that is $\phi_0^h \left( x \right)$, is identically
zero since the BCs are homogeneous and there is no need for an excess function.
The $\mathcal{O} \left( \varepsilon \right)$ equation becomes
\begin{equation*}
  \partial_z^2 \phi_1
  = - \frac{k^2}{m^2} \cos \left( kx \right) \cos \left( mz \right)
\end{equation*}
whose solution is
\begin{equation*}
  \phi_1 \left( x,z \right)
  = -\frac{k^2}{m^2} \frac{\cos \left( kx \right) \cos \left( mz \right)}{-m^2}.
\end{equation*}
Continuing in this manner, we see that the solution at $\mathcal{O} \left(
\varepsilon^n \right)$ is
\begin{equation*}
  \phi_n \left( x, z \right)
  = \left( -\frac{k^2}{m^2} \right)^n
  \frac{ \cos \left( kx \right) \cos \left( mz \right)}{-m^2}
\end{equation*}
which means that if we terminate the leptic solver at $\mathcal{O} \left(
\varepsilon^p \right)$, the solution we arrive at is given by the sum
\begin{equation}
  \label{leptic_series_sol}
  \phi \left( x,z \right)
  = \frac{\cos \left( kx \right) \cos \left( mz \right)}{-m^2}
  \sum_{n=0}^p \left( -\frac{k^2}{m^2} \right)^n
\end{equation}
where we have set $\varepsilon$ to $1$. If $k^2/m^2 \geqslant 1$, then
this geometric series will diverge as $p \rightarrow \infty$ and the leptic
solver will ultimately fail. On the other hand, if $k^2/m^2 < 1$, then
the series will be finite for all values of $p$ and we can put the solution in a
closed form,
\begin{equation*}
  \phi \left( x,z \right)
  = \frac{\cos \left( kx \right) \cos \left( mz \right)}
  {- \left( k^2 + m^2 \right)}
  \left\{ 1 - \left( - \frac{k^2}{m^2} \right)^{p+1} \right\}.
\end{equation*}
In the limit $p \rightarrow \infty$, the term in braces tends to $1$ and we
recover the exact, analytic solution of the elliptic equation.

Since the wavenumbers $k$ and $m$ are both positive, we can simply say that
convergence of the leptic method requires $\max \left( k/m \right) < 1$.
Analytically, this quantity depends on the harmonic content of the source, $\rho
\left( x,z \right)$, and is, in principle, unbounded. Numerically, once a
discretization has been chosen, $k$ and $m$ are limited to a finite number of
values. If we let the source term be a general linear combination of
eigenfunctions and if our rectangular domain has dimensions $L$ by $H$ divided
uniformly into elements of size $\Delta x$ by $\Delta z$, and it is discretized
with a spectral method, then $\max \left( k \right) = \pi/\Delta x$ and $\min
\left( m \right) = \pi/H$. This produces our convergence condition, $H/\Delta x
< 1$, which is the origin of the leptic ratio, $\lambda = \min \left( \Delta x^m
/H \right)$, and its square inverse, $\varepsilon = \max \left( H/\Delta x^m
\right)^2$, used throughout this paper.

Notice that this convergence condition is a restriction on how we must
discretize a given domain, it is not directly a restriction on the source term
of the elliptic equation at hand. This means that the leptic method should
converge similarly for all equations that use a particular uniform, rectangular
grid. If the grid is not rectangular or uniform, then the relavant convergence
condition is $H_i/\Delta x_i < 1$ at all grid positions, $i$. Here, $H_i$ is the
vertical height of the domain at $x_i$ and $\Delta x_i$ is the minimum
horizontal grid spacing at $x_i$.

\subsection{General case}

Now, we will extend this argument to the more general case involving the
positive-definite, symmetric tensor field, $\sigma^{i j} \left( x, z \right)$.
We wish to perform a convergence analysis on
\begin{equation}
  \label{poisson_norot}
  \left\{ \partial_z \sigma^{zz} \partial_z
  + \varepsilon \left( \partial_x \sigma^{xx} \partial_x
  + \partial_x \sigma^{xz} \partial_z
  + \partial_z \sigma^{zx} \partial_x \right) \right\} \phi \left( x,z \right)
  = \rho \left( x,z \right).
\end{equation}

Without the exact form of each $\sigma^{ij} \left( x,z \right)$, we cannot
trivially diagonalize the operator in $(k, m)$-space. We can, however,
diagonalize the operator in $(x,z)$-space by performing a small rotation by an
angle $\frac{1}{2} \tan^{-1} \left( \frac{2 \varepsilon \sigma^{xz}}{\sigma^{zz}
- \varepsilon \sigma^{xx}} \right)$. This casts the equation into the simpler
form
\begin{equation*}
  \left\{ \partial_z \left[ \sigma^{zz}
  + \mathcal{O} \left( \varepsilon^2 \right) \right] \partial_z
  + \partial_x \left[ \varepsilon \sigma^{xx}
  + \mathcal{O} \left( \varepsilon^2 \right) \right] \partial_x \right\}
  \phi \left( x,z \right)
  = \rho \left( x,z \right),
\end{equation*}
where the $x$ and $z$ now represent the new coordinates. We might as well just
let $\sigma^{zz} + \mathcal{O} \left( \varepsilon^2 \right) \rightarrow
\sigma^{zz}$ and $\varepsilon \sigma^{xx} +\mathcal{O} \left( \varepsilon^2
\right) \rightarrow \varepsilon \sigma^{xx}$ so that
\begin{equation}
  \label{poisson_rot}
  \left\{ \partial_z \sigma^{zz} \partial_z
  + \varepsilon \partial_x \sigma^{xx} \partial_x \right\}
  \phi \left( x,z \right)
  = \rho \left( x,z \right).
\end{equation}

Even though each term of eq. (\ref{poisson_rot}) is functionally different than
the corresponding terms of eq. (\ref{poisson_leptic}), their magnitudes are
equal. This means that a convergence analysis of eq. (\ref{poisson_rot}) must
lead to a restriction of the form $\lambda > \mathcal{O} \left( 1 \right)$.
Rotating back to eq. (\ref{poisson_norot}) cannot possibly change this
restriction due to the vanishing size of the rotation angle. This shows that
even in the general case of eq. (\ref{Poisson}), the leptic solver will converge
as long as the discretization is chosen to satisfy $\lambda > \mathcal{O} \left(
1 \right)$.

\subsection{The leptic solver as a preconditioner}\label{sec_leptic_precond}

Let us return to the simple case of solving $\nabla^2 \phi = \cos \left( kx
\right) \cos \left( mz \right)$ on a rectangular domain. After applying $n_l$
iterations of the leptic solver, the residual, $r$, is found via eq.
(\ref{leptic_series_sol}) with $p = n_l - 1$.
\begin{eqnarray*}
  r
  & = & \cos \left( kx \right) \cos \left( mz \right)
  - \nabla^2 \left\{
  \frac{\cos \left( kx \right) \cos \left( mz \right)}{-m^2}
  \sum_{n = 0}^{n_l - 1} \left( - \frac{k^2}{m^2} \right)^n \right\}
  \\
  & = & \cos \left( kx \right) \cos \left( mz \right)
  - \frac{k^2 + m^2}{m^2} \cos \left( kx \right) \cos \left( mz \right)
  \sum_{n = 0}^{n_l - 1} \left( - \frac{k^2}{m^2} \right)^n
  \\
  & = & \left\{ 1
  + \sum_{n = 1}^{n_l} \left( - \frac{k^2}{m^2} \right)^n
  - \sum_{n = 0}^{n_l - 1} \left( - \frac{k^2}{m^2} \right)^n \right\}
  \cos \left( kx \right) \cos \left( mz \right)
  \\
  & = & \left( - \frac{k^2}{m^2} \right)^{n_l}
  \cos \left( k x \right) \cos \left( m z \right)
\end{eqnarray*}
The last line comes from collapsing the telescoping set of sums. This gives us
an amplification factor for each eigenmode of the residual. That is, if we are
given a generic residual and perform an eigenvector expansion,
\begin{equation*}
  r \left( x,z \right)
  = \sum_k \sum_m r \left( k,m \right)
  \cos \left( kx \right) \cos \left( mz \right),
\end{equation*}
then the magnitudes of the individual components, $r \left( k,m \right)$, will
be amplified or attenuated by $k^2/m^2$ each time we iterate. If the grid is
constructed such that $\lambda > \mathcal{O} \left( 1 \right)$, then $r \left(
k,m \right)$ will always be attenuated since $\max \left( k^2/m^2 \right) < 1$.
If, however, the grid's leptic ratio is $\sim \mathcal{O} \left( 1 \right)$,
then only those eigenmodes with $k^2/m^2 < 1$ will diminish and those with
$k^2/m^2 > 1$ will be amplified. In $(x,z)$-space, this effect appears as a
diverging solution, but in $(k,m)$-space, we can see that the solution is split
into converging and diverging parts -- we are conditioning the solution. For
this reason, even though preconditioning is normally understood as the action of
substituting the original operator with a modified one with better spectral
properties \cite{vanderhorst}, we will use preconditioning to mean the action of
replacing an initial guess with one that has better spectral support.

\begin{figure}[h!]
  \centering
  \includegraphics[width=\textwidth]{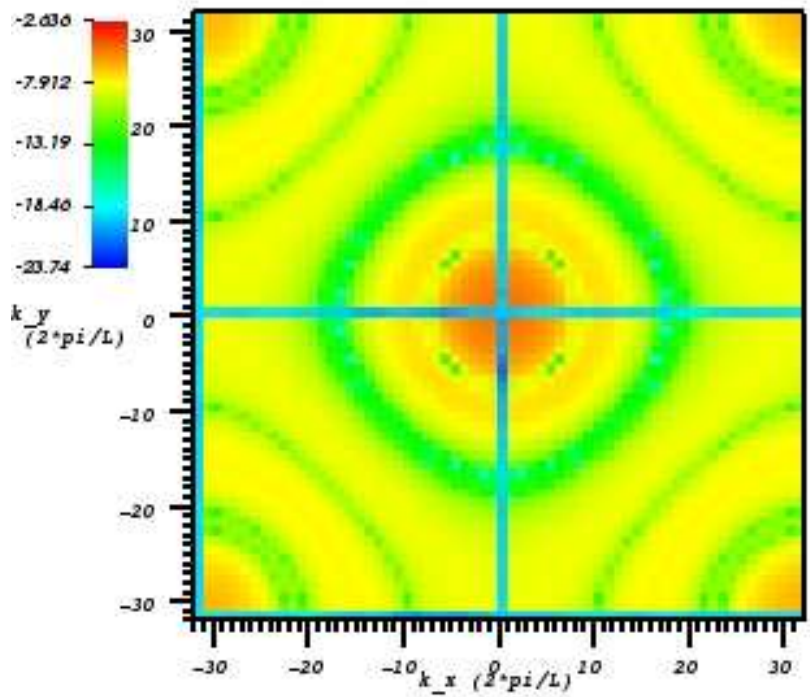}
  \caption{A 2-dimensional slice of the residual error after 26 iterations of
BiCGStab. Colors are on a logarithmic scale. Notice the large error near the
center of the plot, indicating BiCGStab's difficulty eliminating low frequency
errors. The blue lines are the zero-frequency modes that must be fixed to agree
with the boundary conditions.}
  \label{fig_bicgstab}
\end{figure}

\begin{figure}[h!]
  \centering
  \includegraphics[width=\textwidth]{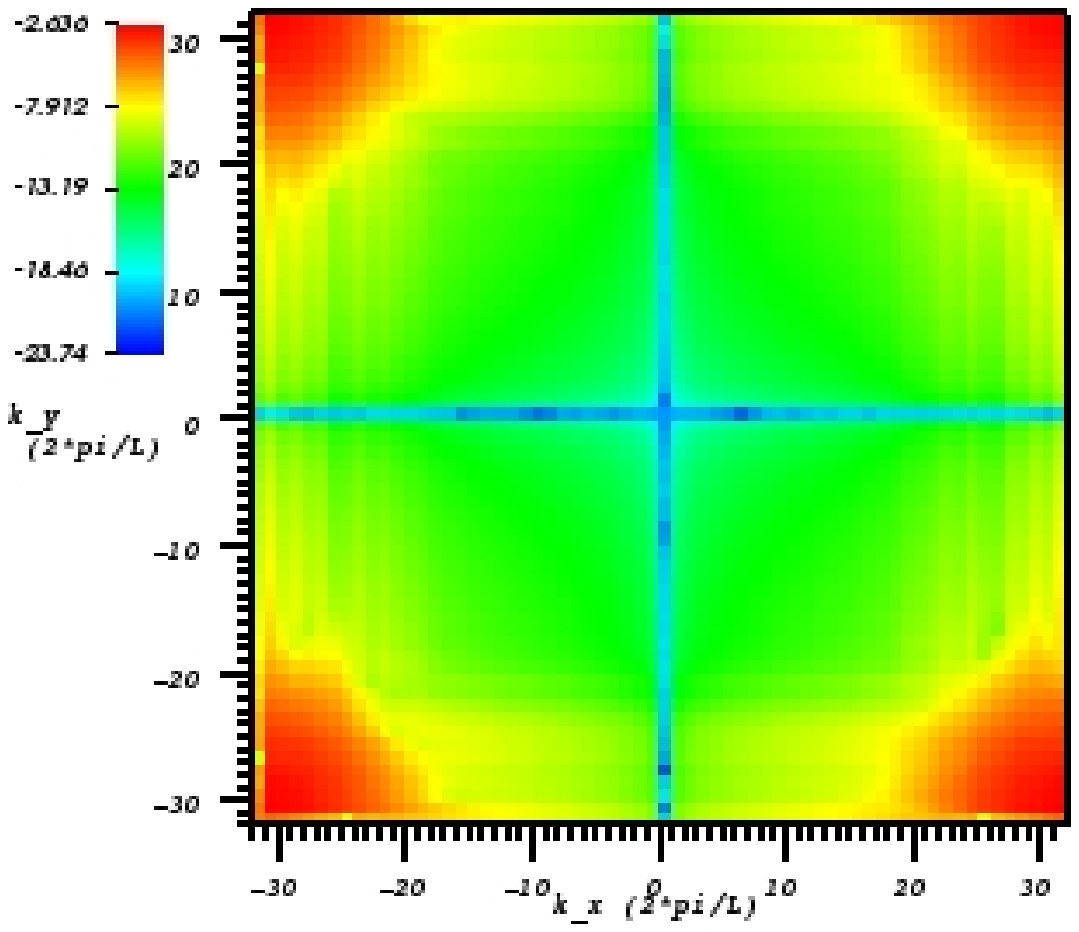}
  \caption{The residual error after 24 iterations of the leptic solver. This
solver eliminates low frequency errors much more effectively than high frequency
errors, indicating the leptic solver's potential to serve as a preconditioner
for BiCGStab.}
  \label{fig_leptic}
\end{figure}

As an example, consider a $64 \times 64 \times 16$ grid with $\Delta x = \left(
1,1,0.1 \right)$. This fixes $\varepsilon$ at $2.56$, which is large
enough to cause problems for the leptic solver.\footnote{For the remainder of
this paper, we will be dealing with a geometric anisotropy quantified by
$\lambda$. The perturbation parameter is then $\varepsilon = \left( H/\Delta x
\right)^2$.} In order to learn how the solvers are treating the modes on this
grid, let's apply them to
\begin{equation*}
  \nabla^2 \phi
  = \sum_{i = 1}^{32} \sum_{j = 1}^{32} \sum_{k = 1}^8
  \cos \left( \frac{2 \pi i x}{L} \right)
  \cos \left( \frac{2 \pi j y}{L} \right)
  \cos \left( \frac{2 \pi k z}{H} \right)
\end{equation*}
with homogeneous boundary conditions. This is a residual equation whose r.h.s.
harbors every periodic mode supported by the grid in equal amounts (except for
the zero modes, which must be removed to be consistent with the boundary
conditions). We only consider periodic modes to facilitate spectral analysis via FFT. In one test, we solved this equation with a BiCGStab solver and in another
separate test, we used the leptic method. BiCGStab stalled in 26 iterations and
the leptic solver began to diverge after 24 iterations. Once progress came to a
halt, we performed an FFT to locate which modes were converging slowest. To
simplify the visualization, we found the $(k_x,k_y)$ slice that contained the
most slowly converging modes (which, consistent with the previous analysis, is
the smallest value $k_z = 2 \pi/H$). The results are in figures
\ref{fig_bicgstab} and \ref{fig_leptic}.\footnote{VisIt has been developed and
is being distributed by the Lawrence Livermore National Laboratory. Website:
https://wci.llnl.gov/codes/visit.}

\begin{figure}[h!]
  \centering
  \includegraphics[width=\textwidth]{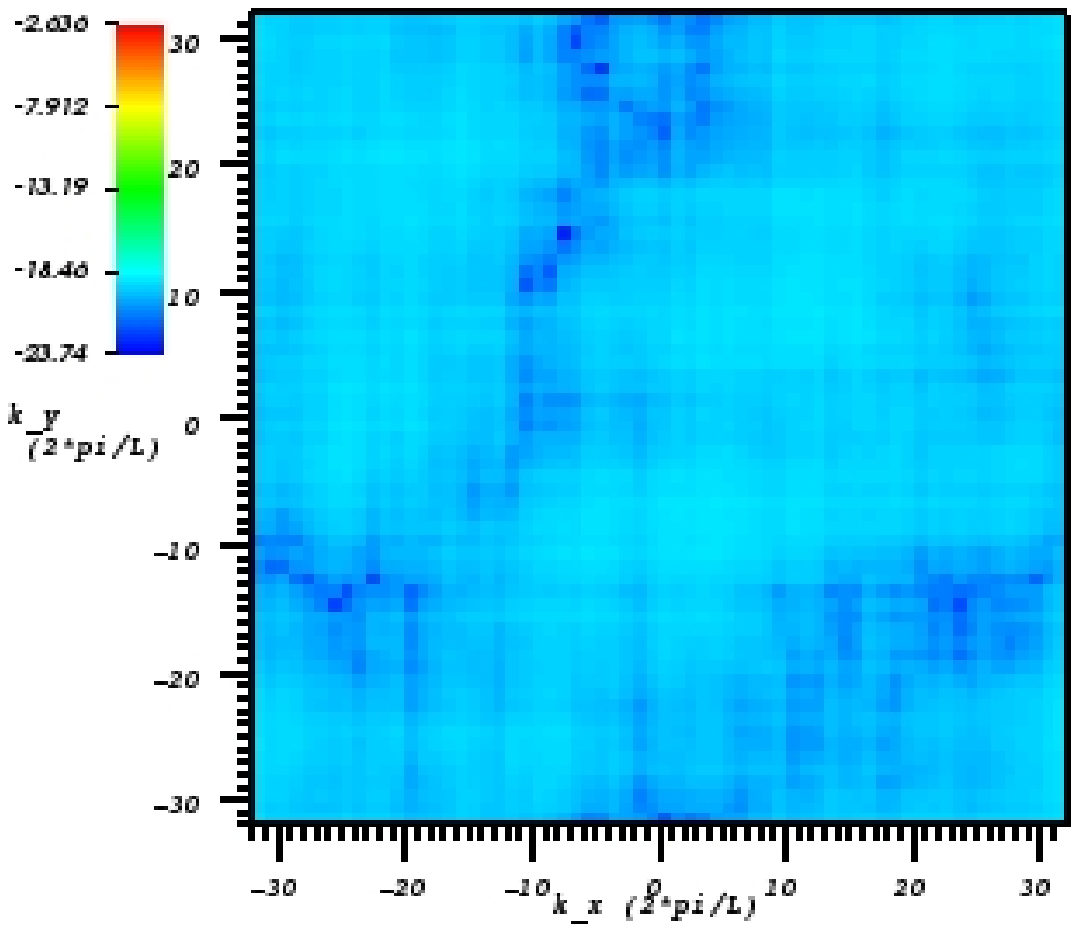}
  \caption{The residual error of the BiCGStab method when given an initial guess
generated by the leptic solver.}
  \label{ref_leptic_bicgstab}
\end{figure}

The color in these plots represent the base-10 logarithm of the Fourier
coefficients of each mode. It is apparent that each method has its own distinct
problem region shown in red. The BiCGStab solver has the most trouble dealing
with low frequency modes while the leptic solver has trouble with high frequency
modes. Since the leptic solver produced a residual whose largest modes can
easily be handled by BiCGStab, we apply the BiCGStab using as initial guess the
output of the leptic solver after 18 iterations. Now BiCGStab is able to
converge quickly (figure \ref{ref_leptic_bicgstab}).

We should mention that this is just an illustration. In this example, the
BiCGStab method on such a small grid would have converged on its own in a
reasonable number of iterations. On a larger grid, BiCGStab alone often
converges too slowly to be a viable solution method and sometimes stalls due to
the condition number of the operator. Further complications arise when we use
mapped coordinates because this tends to drive the condition number of the
operator even higher. In these situations, using the leptic method to generate a
suitable initial guess becomes quite useful. We will illustrate this effect in
further detail in section \ref{sec_dem}.

\section{Demonstrations}\label{sec_dem}

In this section, we will create a sample problem on various numerical domains in
order to compare the effectiveness of traditional solvers with methods that
utilize the leptic solver. Our traditional solver of choice will be the BiCGStab
method preconditioned with the incomplete Cholesky factorization (IC) of the
elliptic operator \cite{TeranishiR07}. For simplicity, we will use a rectangular
domain and the r.h.s. will be generated by taking the divergence of a vector
field
\begin{eqnarray*}
  u^{\star 1}
  & = & \left( \frac{z}{L_z} \right)^2 \sin \left( \frac{\pi y}{L_y} \right)
  + \frac{x}{\sqrt{2} L_z}
  \\
  u^{\star 2}
  & = & \left( \frac{z}{L_z} \right)^2 \sin \left( \frac{\pi x}{L_x} \right)
  \\
  u^{\star 3}
  & = & - \left( \frac{z}{L_z} \right)^2 \cos \left( \frac{\pi z}{4 L_z} \right)
  \\
  \rho
  & = & \partial_i u^{\star i}.
\end{eqnarray*}
This vector field also generates the boundary conditions. To compare the
solvers, we will plot the relative residual as a function of the iteration
number. For what follows, the vertical and horizontal solves of the leptic
solver will each be counted each as an iteration.

\subsection{\label{Highleptcart}High leptic ratio - Cartesian coordinates}

\begin{figure}[h!]
  \centering
  \includegraphics[width=\textwidth]{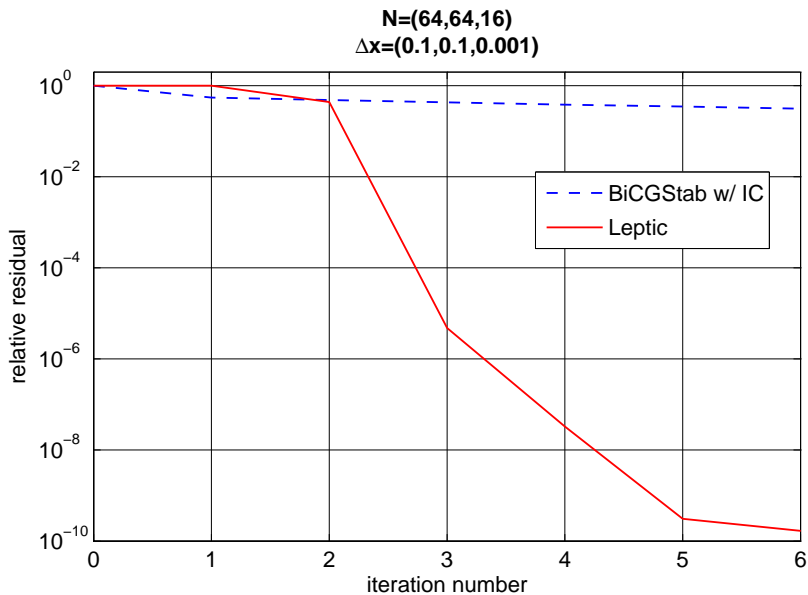}
  \caption{With $\varepsilon \approx 1/40$, the leptic solver is clearly
more efficient than BiCGStab. Since we are using Cartesian coordinates, the
leptic solver only needed to perform one horizontal solve.}
  \label{fig_high}
\end{figure}

First, we set $N = \left( 64,64,16 \right)$ and $\Delta x = \left(
0.1,0.1,0.001 \right)$. This fixes $\varepsilon$ at 0.0256, which lies well
within the region where the leptic solver outperforms traditional methods.
Figure \ref{fig_high} shows the results. The leptic solver is clearly the more
efficient method. In only $6$ iterations, it is able to achieve a relative
residual of $~10^{-10}$. We would have needed $170$ iterations of the
BiCGStab/IC solver to obtain a residual error of that magnitude.

\subsection{Borderline cases - Cartesian coordinates}

\begin{figure}[h!]
  \centering
  \includegraphics[width=\textwidth]{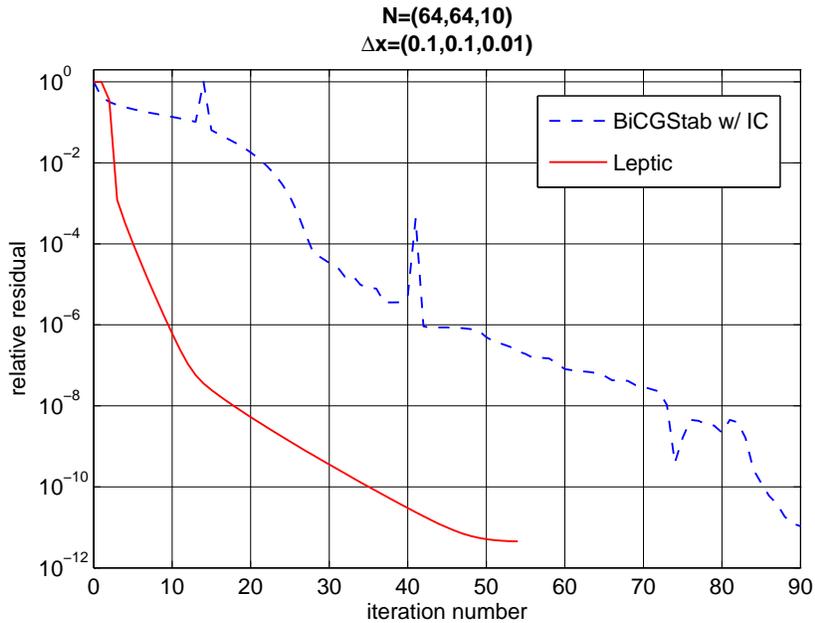}
  \caption{The convergence patterns of the leptic and BiCGStab solvers when
  $\varepsilon = 1$ and condition number $\approx 10^{5.6}$. The spikes in the
  BiCGStab residual are due to restarts.}
  \label{fig_64_64_10}
\end{figure}

When $\varepsilon =\mathcal{O} \left( 1 \right)$, the leptic solver may or may
not be the most efficient solver. We will denote these situations as
{\em borderline cases}. In the first borderline case, we will bring
$\varepsilon$ to $1$ by setting $N = \left( 64,64,10 \right)$ and $\Delta x =
\left( 0.1,0.1,0.01 \right)$. Figure \ref{fig_64_64_10} shows that the leptic
solver requires approximately $5$ times as many iterations as it did in the
$\varepsilon = 0.0256$ example to achieve an $\mathcal{O}\left(10^{-10}\right)$
residual error. The BiCGStab/IC method, however, converges a bit more quickly
than it did in the previous example. It required only 90 iterations to catch up
to the leptic method. This is emperical evidence of our theoretical assertion --
by raising $\varepsilon$ (decreasing the lepticity), the leptic solver becomes
less effective and the traditional method becomes more effective. In this
specific borderline case, the leptic solver outperforms the BiCGStab method.

\begin{figure}[h!]
  \centering
  \includegraphics[width=\textwidth]{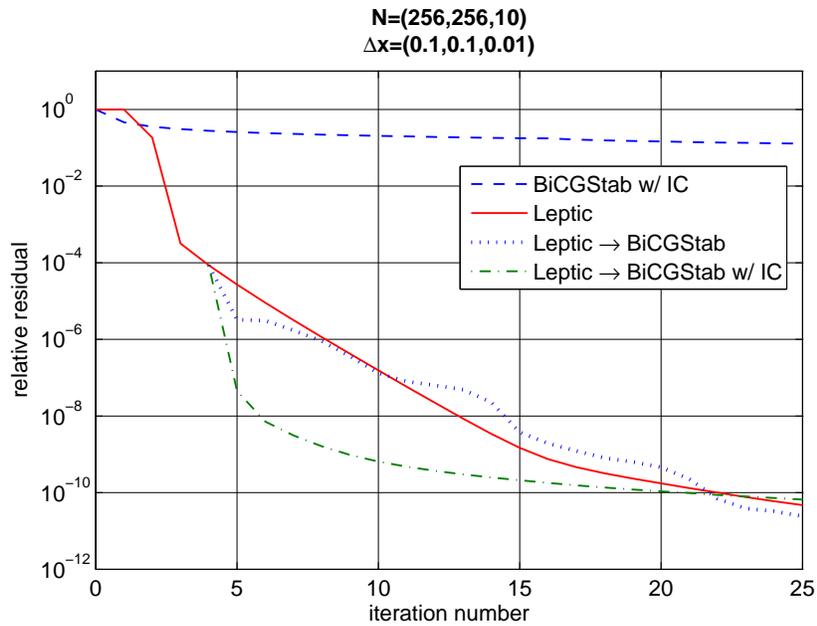}
  \caption{The performance of various solution methods with $\varepsilon = 1$
  and condition number $\approx 10^{6.8}$. After a few iterations of the leptic
  solver, BiCGStab can achieve a fast convergence rate. Using a preconditioner
  such as an incomplete Cholesky decomposition can drive this rate even higher.}
  \label{fig_256_256_10}
\end{figure}

By varying $N_x$ and $N_y$, we can generate an entire class of grids with the
same $\varepsilon$. For example, if we bring $N$ up to $\left(256,256,10\right)$,
the BiCGStab method should not converge as rapidly as before. On the other hand,
$\varepsilon$ is still $1$, which means the leptic solver should perform almost
as well as it did on the $64 \times 64 \times 10$ grid. This is because most of
the leptic solver's convergence relies on the vertical solver. This is true in
general when the horizontal solver is able to be switched off (see section
\ref{sec_elim_horiz}) -- as the horizontal domain grows, the leptic solver
outperforms traditional relaxation methods. This effect is shown in figure
\ref{fig_256_256_10}. By comparing figures \ref{fig_64_64_10} and
\ref{fig_256_256_10}, we see that unlike the leptic method, the BiCGStab/IC
method is in fact slowed down by the larger horizontal grid.

The true value of the leptic solver is illustrated by when we use it to generate
a suitable initial guess for the BiCGStab/IC solver (see section
\ref{sec_leptic_precond}). This initial guess has an error that is dominated by
high wavenumbers. The BiCGStab solver then rapidly removes those errors as shown
by the dash-dot curve in figure \ref{fig_256_256_10}.

\begin{figure}[h!]
  \centering
  \includegraphics[width=\textwidth]{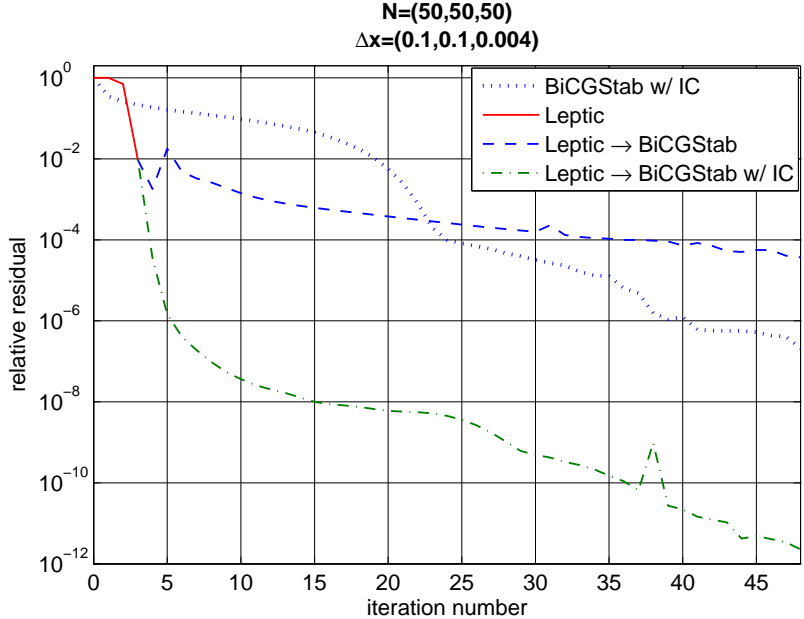}
  \caption{Performance of the solvers with $\varepsilon = 4$ and condition
  number $\approx 10^{6.2}$. The leptic solver began diverging after it's third
  iteration, so control was passed to the Krylov solver. Again, the leptic
  solver proves most valuable as a preconditioner for the BiCGStab/IC solver
  when $\varepsilon = \mathcal{O} \left( 1 \right)$.}
  \label{fig_50_50_50}
\end{figure}

As our final isotropic, borderline case, we will set $N = \left(50,50,50\right)$
and $\Delta x = \left(0.1,0.1,0.004\right)$. This is appropriate for a cubic,
vertically stratified domain and brings $\varepsilon$ up to 4. The BiCGStab/IC
solver does not provide immediate convergence and the leptic method would have
started to diverge after its third iteration, but when we combine the two
methods as we did in the last example, we see a much more rapid convergence than
either method could individually achieve (figure \ref{fig_50_50_50}).

\subsection{High leptic ratio - Mapped coordinates}

\begin{figure}[h!]
  \centering
  \includegraphics[width=\textwidth]{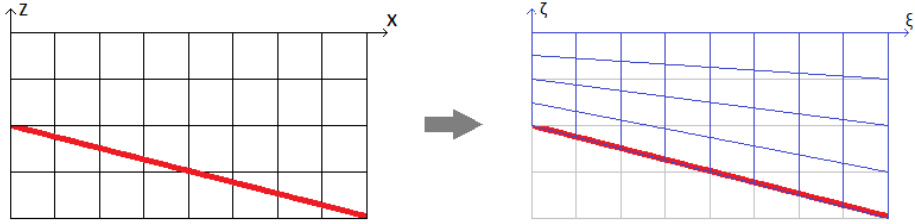}
  \caption{A cross section of the coordinate mapping. The thick line denotes the
  lower vertical boundary.}
  \label{mapping}
\end{figure}

When the metric is diagonal, several of the terms in our expansion (sec.
\ref{Expansion_Summary}) vanish. This means we can remove much of the code to
produce a more efficient algorithm. This reduced code is what generated the
results of the previous sections. However, in these simple geometries the value
of the leptic expansion is somewhat limited because it is normally possible to
employ fast direct solvers. Not so in the case we consider now, where we apply
the full algorithm by considering a non-diagonal metric. The metric we will use
is routinely employed in meteorological and oceanic simulations of flows over
non uniform terrain. It maps the physical domain characterized by a variable
topography $z = h (x,y)$ to a rectangular computational domain. Although any
relief may be specified, it is sufficient for our illustrative purposes to
simply let the depth go from $d/2$ to $d$ linearly as $x$ goes from $0$ to $L$,
where $d < 0$ (Figure \ref{mapping}).

We define $h (\xi) = \frac{d}{2} + \frac{d}{2 L} \xi$, where $\left( x,y,z
\right) \rightarrow \left( \xi, \eta, \frac{h (\xi)}{d} \zeta \right)$. This
means that the symmetric tensor, $\sigma^{i j} = \sqrt{g} g^{i j}$, is given by
\begin{equation*}
  \sqrt{g} g^{ij}
  = \left( \begin{array}{ccc}
    \frac{\xi + L}{2 L} & 0 & - \frac{\zeta}{2 L}\\
    0 & \frac{\xi + L}{2 L} & 0\\
    - \frac{\zeta}{2 L} & 0 & \frac{4 L^2 + \zeta^2}{2 L (\xi + L)}
  \end{array} \right)^{i j}.
\end{equation*}
For the next two examples, we will be seeking a simple
\[ \phi_{\text{sol}} \left( \xi,\eta,\zeta \right)
   = \cos \left( \frac{2 \pi \xi}{L} \right)
     \cos \left( \frac{2 \pi \eta}{L} \right)
     \cos \left( \frac{2 \pi \zeta}{d} \right)
\]
solution by setting $u^{\star i} = \sqrt{g} g^{ij} \partial_j \phi_{\text{sol}}$
and $\rho = \partial_i u^{\star i}$.

Setting $N = \left( 256,256,64 \right)$ and $\Delta x = \left( 0.25,0.25,0.0025
\right)$ gives us $\varepsilon = 0.4096$. After approximately 200 iterations,
the BiCGStab method begins to stall with a relative residual of $\mathcal{O}
\left( 10^{- 8} \right)$. On the other hand, the leptic solver was able to reach
a relative residual of $3.637 \times 10^{- 9}$ in only 10.2\% of the time before
terminating at $\mathcal{O} \left( \varepsilon^4 \right)$. The reason the leptic
method stopped iterating was due to inconsistent data given to the vertical
solver. In section \ref{Derivation}, we showed that at each iteration, a 2D
poisson problem, namely eq. (\ref{phi0h}), must be handed to a traditional
solver and if that solver fails, the next vertical stage may have a source term
that is not compatible with its boundary conditions. Remarkably, the horizontal
stages failed (abandoned due to stalling) at every order of calculation and the
leptic solver was still able to out-perform the BiCGStab method. Had our
decision-making algorithm been altered to abandon the horizontal solver after
failing the first time, which is reasonable for some problems, we would have
converged much faster. It should be pointed out that until $\mathcal{O}
\left( \varepsilon^4 \right)$, the leptic solver did not diverge, therefore we
never needed to transfer control to a full 3D BiCGStab solver -- which would
have been slow.

\begin{figure}[h!]
  \centering
  \includegraphics[width=\textwidth]{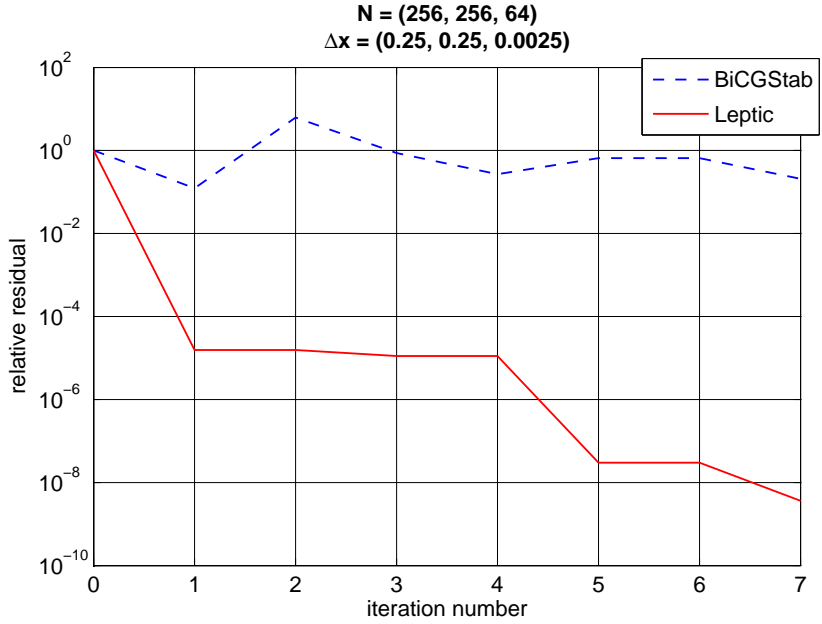}
  \caption{Performance of the leptic and Krylov solvers with an anisotropic
  $\sigma^{ij}$ and $\varepsilon \approx 0.4$. Since our solver is using
  Chombo's matrix-free methods (known as \textit{shell matrices} in some popular
  computing libraries such as PETSc \cite{petsc-web-page}), the Cholesky
  decomposition of the elliptic operator can not be performed.}
  \label{fig_mapped_high}
\end{figure}

\subsection{Borderline case - Mapped coordinates}

\begin{figure}[h!]
  \centering
  \includegraphics[width=\textwidth]{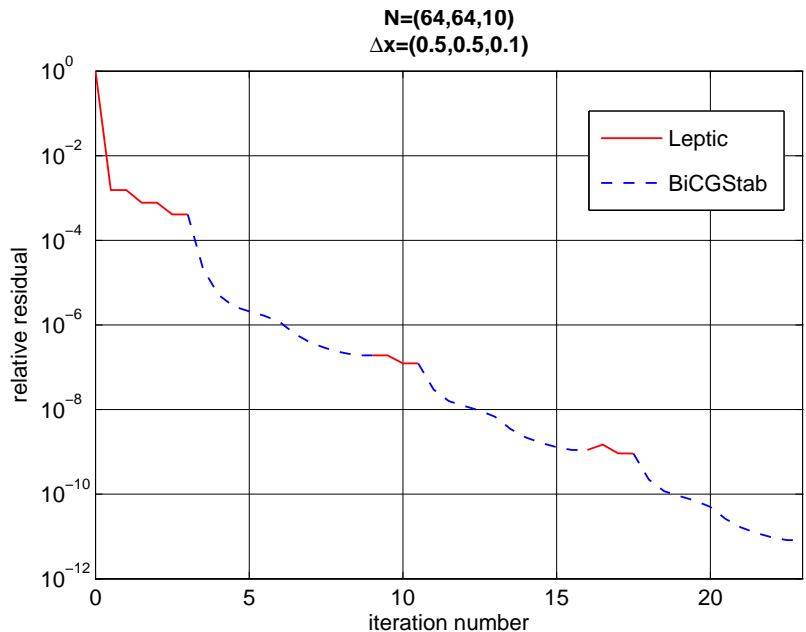}
  \caption{The convergence pattern of a hybrid leptic/Krylov method when
  $\varepsilon = 4$ on an anisotropic domain. In this case, neither method would
  have individually provided rapid convergence, but when the methods are
  combined, we see a very rapid convergence and the introduction of another
  preconditioner (eg. IC) is not necessary.}
  \label{fig_mapped_borderline}
\end{figure}

For this scenario, we set $N = \left( 64,64,10 \right)$ and $\Delta x = \left(
0.5,0.5,0.1 \right)$ giving us $\varepsilon = 4$. To solve our sample problem on
this grid, we let the leptic and BiCGStab solvers work together, iteratively. As
soon as one solver begins to stall or diverge, control is passed to the other
solver. The effectiveness of this algorithm is explained in section
\ref{sec_leptic_precond} -- the leptic solver first reduces low frequency errors
until high frequency errors begin to dominate the residual, then the BiCGStab
solver reduces high frequency errors until low frequency errors dominate the
error. This continues until the residual error has been reduced over the entire
spectrum supported by the grid. Figure \ref{fig_mapped_borderline} illustrates
the effectiveness of this algorithm rather convincingly.

\section{Discussion}\label{discussion}

The leptic method was originally devised as a method to add a physically
appropriate amount of dispersion when numerically modeling the propagation of
nonlinear waves in a dispersive medium. In this paper, we generalized the method
so it could be used to actually speed up the numerical solution of Poisson
problems characterised by a high level of anisotropy. The key idea is that
instead of (or in addition to) preconditioning the operator to achieve overall
better spectral properties, we precondition the initial guess (or the restarts)
to achieve better spectral support of the residual by coupling the leptic
expansion to Krylov methods. However, since the former converges on
its own if the lepticity of the grid is sufficiently large, it is easy to see
that it could be used within multigrid methods as well. Namely, when the
coarsening reaches a point that the lepticity of the grid is below critical, the
leptic expansion can be used in lieu of the relaxing stage to generate an exact
solution at the coarse level. This would likely cut down the layers of
coarsening.    

In its full generality, this method can be used to solve anisotropic Poisson
equations in arbitrary, $D$-dimensional coordinate systems. In many cases,
however, numerical analysis is performed using simple, rectangular coordinates
without an anisotropic tensor, $\sigma^{ij}$. This simplification reduces much
of the computation. For these common purposes, we included a summary of the
Cartesian version of the expansion. Both the general and Cartesian expansions
are summarized in section \ref{Expansion_Summary}.

When implementing the leptic method for numerical work, the computational domain
should be split in all but the vertical (stiff) dimension. The vertical ordinary
differential equations require a set of integrations -- one for each point in
the horizontal plane. With this domain decomposition, these integrations are
independent of one another and the solutions to the vertical problems can be
found via embarassingly parallel methods. On the other hand, the solutions to
the horizontal equations cannot be parallelized as trivially, making their
solutions more costly to arrive at. In light of this, section
\ref{sec_elim_horiz} was provided to discuss when it is appropriate to eliminate
these expensive stages.

The computational cost of a single iteration is of the same order as a
preconditioned step of a Krylov method. The savings are obtained in the faster
rate of convergence, shown in the examples above over a wide range of
anisotropic conditions, as well as the relative ease with which the leptic
expansion can be parallelized. The best rate of convergence is achieved by using
the leptic expansion at the beginning and every time the convergence rate of the
Krylov method slows down. We did not attempt  to predict a priori after how many
steps the switch is necessary. If the Poisson problem, as it often happens, is
part of a larger problem that is solved many times, a mockup problem should be
solved at the beginning to determine empirically the best switch pattern.

Adapting the expansion to accept Dirichlet boundary conditions would require a
new derivation similar to that of section \ref{Derivation}, but the job would be
much simpler since Dirichlet elliptic operators have a trivial null space,
thereby eliminating the need to consider compatibility conditions among the
second-order operators and their boundary conditions.

{\bf Acknowledgment:} This work was supported by grants from ONR and NSF.
\newpage

\bibliographystyle{model1-num-names}
\bibliography{tpb}

\end{document}